# Spin dynamics of edge-sharing spin chains in $SrCa_{13}Cu_{24}O_{41}$

Guochu Deng[a*], Dehong Yu[a], Richard Mole[a], Ekaterina Pomjakushina[b], Kazimierz Conder[b], Michel Kenzelmann[c], Shin-ichiro Yano[d], Chin-Wei Wang[d], Kirrily C. Rule[a], Jason S. Gardner[d], Huiqian Luo[e], Shiliang Li[e], Clemens Ulrich[f], Paolo Imperia[a], Wei Ren[g], Shixun Cao[g], Garry J. McIntyre[a*]

[a] *Australian Centre for Neutron Scattering, Australian Nuclear Science and Technology Organisation, New Illawarra Road, Lucas Heights NSW 2234, Australia*

[b] *Laboratory for Multiscale Materials Experiments, Paul Scherrer Institut, CH-5232 Villigen-PSI, Switzerland*

[c] *Laboratory for Neutron Scattering and Imaging, Paul Scherrer Institut, CH-5232 Villigen-PSI, Switzerland*

[d] *National Synchrotron Radiation Research Centre, Hsinchu 30076, Taiwan*

[e] *Beijing National Laboratory for Condensed Matter Physics, Institute of Physics, Chinese Academy of Sciences, Beijing 100190, China*

[f] *School of Physics, University of New South Wales, Kensington NSW 2052, Australia.*

[g] *Department of Physics, International Centre of Quantum and Molecular Structures, and Materials Genome Institute, Shanghai University, Shanghai 200444, China*

## Abstract

The low-energy magnetic excitation from the highly Ca-doped quasi-one-dimensional magnet $SrCa_{13}Cu_{24}O_{41}$ was studied in the magnetic ordered state by using inelastic neutron scattering. We observed the gapless spin-wave excitation, dispersive along the *a* and *c* axes but nondispersive along the *b* axis. Such excitations are attributed to the spin wave from the spin-chain sublattice. Model fitting to the experimental data gives the nearest-neighbour interaction $J_c$ as 5.4 meV and the interchain interaction $J_a$ = 4.4 meV. $J_c$ is antiferromagnetic and its value is close to the nearest-neighbour interactions of the similar edge-sharing spin-chain systems such as $CuGeO_3$. Comparing with the hole-doped spin chains in $Sr_{14}Cu_{24}O_{41}$, which shows a spin gap due to spin dimers formed around Zhang-Rice singlets, the chains in $SrCa_{13}Cu_{24}O_{41}$ show a gapless excitation in this study. We ascribe such a change from gapped to gapless excitations to holes transferring away from the chain sublattice into the ladder sublattice upon Ca doping.

E-mails: guochu.deng@ansto.gov.au, garry.mcintyre@ansto.gov.au

**Introduction**

The dimensionality of a spin system is one of the most important characters for its underlying physics. Most of the three-dimensional (3D) Heisenberg magnetic systems can be well described by the linear spin-wave theory (LSWT), which was developed decades ago[1]. Upon reducing a magnetic system from 3D to two or one dimensions (2D or 1D), quantum fluctuations will be substantially enhanced, which consequently results in various intriguing and exotic physical phenomena.[2] The theoretical work by Bethe more than 80 years ago predicted that the S = ½ antiferromagnetic Heisenberg chain cannot form a long-range magnetic ordering due to strong quantum fluctuations.[3] A domain-wall-like spin excitation called "spinon" was proposed later, which has been experimentally confirmed in various real magnetic systems. Unfortunately, except for the S = ½ spin-chain case, no exact analytical solution like Bethe's work has been developed for other 1D antiferromagnetic Heisenberg spin systems. However, numerical methods were widely employed to study the ground states of more complicated 1D systems,[4] such as 1D spin chains with S > ½ and spin ladders with even/odd legs[5], and successfully predicted their ground state and dynamic behaviours.

As the simplest case of a 1D spin system, $Cu^{2+}$ (S = ½) spin chains are extensively investigated due to their close relationship with the cuprate high-temperature superconductors.[5] $CuO_2$ chains can be divided into two classes, corner-sharing spin chains and the edge-sharing spin chains. The former are building blocks of cuprate superconductors and spin ladders. In corner-sharing $CuO_2$ chains, the exchange interaction between the nearest neighbours (NN) comes from the super-exchange through the ~180° Cu-O-Cu pathway.[6] Such an interaction is experimentally demonstrated to be quite strong, ranging from 100 meV to 160 meV.[7] In contrast, the NN exchange interaction along edge-sharing spin chains is much weaker due to the ~90° Cu-O-Cu pathway.[8-10] The limited examples of edge-sharing cuprates share fewer common features due to the sensitivity of the interaction on the Cu-O-Cu bond angle.[10] Both ferromagnetic and antiferromagnetic interactions have been reported for this case.[9, 11-13]

$Sr_{14-x}Ca_xCu_{24}O_{41}$ is a quasi-1D S = ½ magnetic system, involving both corner-sharing and edge-sharing $CuO_2$ spin chains.[14] The corner-sharing spin chains form two-leg spin ladders, as shown in Fig. 1(a). Therefore, this series of compounds can be considered to be a very good playground for studying the physics of these two different 1D/quasi-1D magnetic systems. The other two special properties of this series make it more attractive to physicists. Firstly, this compound becomes superconductive under external pressure and by substituting Ca on the Sr site.[15] Bearing in mind though, it has no $CuO_4$ square plackets in this compound like other superconducting cuprates, but spin chains and spin ladders. Secondly, this system is intrinsically doped with holes, which can be driven back and forth between the chains and the ladders by simply controlling the doping level of Ca.[14] Therefore, one is not only able to study the superconducting mechanism of the spin ladder, but also able to investigate the physics of hole-doped or undoped spin chains/ladders simply by controlling the Ca content. Furthermore, the relationships or interactions between these phenomena are more complicated topics to be addressed.

A recent study by some of us showed how the incommensurate structures of the $Sr_{14-x}Ca_xCu_{24}O_{41}$ series evolve with the Ca doping content, which provides direct structural evidence for the charge transfer from the spin chains to the ladders by increasing Ca doping level.[14] Our inelastic neutron scattering experiments revealed that the increase of Ca doping did not close the spin gap of the spin ladder in all these compounds, but only suppressed its intensity due to the increase of hole doping.[16] In the highly doped compound $SrCa_{13}Cu_{24}O_{41}$, long-range magnetic ordering has been observed and the coexistence of magnetic ordering and a spin-liquid ground state was demonstrated.[17] All these studies provide some answers to the interesting questions mentioned above. However, many open questions still remain.

In this study, we investigate the low-energy excitation from the magnetic ordered phase in $SrCa_{13}Cu_{24}O_{41}$ by inelastic neutron scattering. Considering the hole transfer from the chains to the ladder sublattice, this work is a comparative research to the study of the hole-doped spin chain in the parent compound $Sr_{14}Cu_{24}O_{41}$. In contrast to the singlet ground state of $Sr_{14}Cu_{24}O_{41}$, the observed spin-wave excitation in $SrCa_{13}Cu_{24}O_{41}$ demonstrates its 2D feature rather than a 1D one in this compound. The physics of these two scenarios have been compared. The difference between the current compound and other edge-sharing cuprate systems has been addressed in order to provide a more general view of this special spin-chain system.

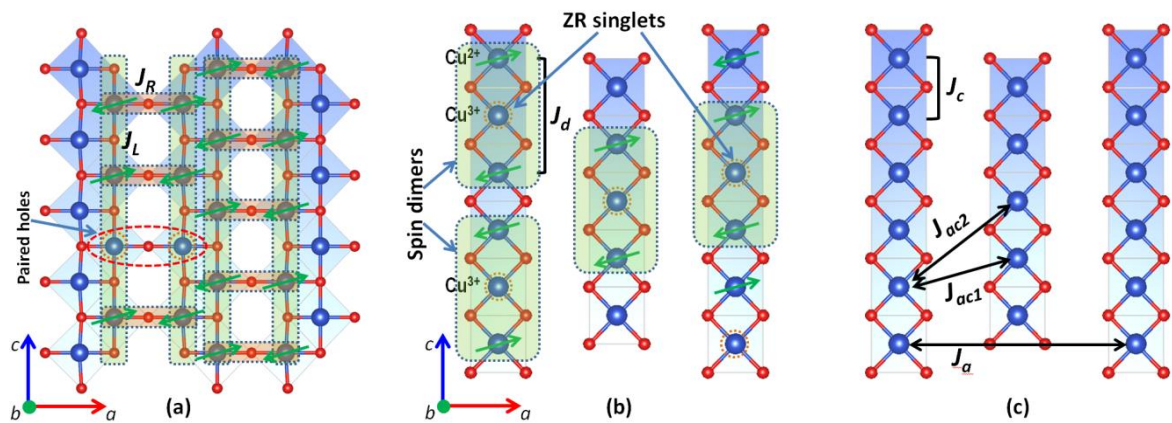

Fig. 1 (a) the spin-ladder sublattice of $Sr_{14-x}Ca_xCu_{24}O_{41}$ consists of two-leg ladders, $J_R$ and $J_L$ are exchange interactions along the rungs and the legs; Theory predicts that doped holes form pairs (see the red ellipse), which play important roles in its superconducting phase. (b) On the spin-chain sublattice of $Sr_{14}Cu_{24}O_{41}$, so-called Zhang-Rice (ZR) singlets are formed by a copper hole hybridized with a hole on its neighbouring oxygen. The two neighbouring $Cu^{2+}$ form spin dimers through the nonmagnetic $Cu^{3+}$.[18] (c) The holes in highly Ca-doped compounds were primarily driven to the spin-ladder sublattice due to the chemical pressure, which consequently leaves the spin chain nearly undoped. This magnetic system becomes long-range magnetic order at 4.2K. $J_c$ is the NN exchange interaction while $J_{ac1}$, $J_{ac2}$ and $J_a$ are possible interactions along the $a$ axis.

## Spin Ladder and Chain Physics of $Sr_{14-x}Ca_xCu_{24}O_{41}$

The $Sr_{14-x}Ca_xCu_{24}O_{41}$ series is formed by stacking the spin-ladder (Fig. 1(a)) and spin-chain (Fig. 1(b) and (c)) sublattices along the $b$ axis.[14] The ladder units and spin chain units are incommensurate along the $c$ axis. These two sublattices demonstrate quite different physics, such as spin dimerization,[19] superconductivity,[15] charge crystallization,[20] charge transfer,[14] and so on, which are remarkably changed by Ca doping, too. Here we will briefly introduce them separately in support of the further discussion in this study.

### (A) Spin Ladder

According to the theory,[5] even-leg half-spin ladders are gapped spin systems no matter the ratio of $J_R/J_L$, where $J_R$ is the interaction along the rungs and $J_L$ is the interaction along the legs (see Fig. 1(a)). In the case of $J_R/J_L \gg 1$, the spins on each rung form a singlet dimer, which is the ground state of an even-leg spin-ladder system. As proposed by Dagotto and Rice,[5] by lightly doping an S = ½ even-leg spin-ladder system with holes, one could expect $d$-wave superconductivity with a spin gap in this ladder due to hole pairing along the rungs, as shown in Fig. 1(a). Such superconductivity is very similar to what is observed in 2D cuprates. Both of them possess a spin gap and $d_{x^2+y^2}$ symmetry. From this point of view, disclosing the superconducting mechanism in spin ladders undoubtedly helps to improve our understanding of the superconductivity in 2D cuprates.[5]

The spin ladder in $Sr_{14-x}Ca_xCu_{24}O_{41}$ attracts much more attention than the spin chain because the former was suggested to be the platform of the observed superconductivity in the highly Ca-doped compounds.[21] In the parent compound, $Sr_{14}Cu_{24}O_{41}$, a strong spin-gap excitation was observed by using inelastic neutron scattering,[16, 22] indicating $J_L > J_R$ in the parent compound. Correspondingly, no superconductivity was observed under external pressure up to 6.5 GPa,[23] in contrast to the superconductivity of highly Ca-doped compounds at ~ 3 GPa[23] or even lower uniaxial pressure[24]. So far, there is no solid theoretical understanding why the isovalent doping with Ca changes the superconductivity of this series of compounds.

Ca-doping extensively changes the physics of the ladder sublattice, which may induce superconductivity theoretically predicted. First of all, holes transfer to the ladder sublattice, which makes hole pairing possible as described in the theory. Secondly, we confirmed that the spin gap, which is the driving force for the hole pairing, still exists at ~ 32.5 meV in highly Ca-doped compounds.[17] The high pressure could play another critical role in changing the ratio $J_R/J_L$ from less than 1 to larger than 1 by distorting the lattice even though our previous measurement of the spin gap under a hydrostatic pressure of 2 GPa was inconclusive due to the challenges of the experiments.[25] These facts indicate that Ca-doping play critical role in the superconductivity in this spin ladder system, just as the theory predicted.

### (B) Spin Chain

The spin-chain sublattice of $Sr_{14-x}Ca_xCu_{24}O_{41}$ consists of edge-sharing $CuO_2$ squares, as shown in Fig. 1 (b) and (c), which are separated from the spin ladder by the layer of alkaline-earth cations.[14] In the parent compound, the spin chains look like wrinkled ribbons running along the $c$ axis, in which the Cu-O bond length varies over a quite large range due to the incommensurate modulation of the crystal structure. Such a modulation is strongly suppressed in the highly Ca-doped compounds. The wrinkles on the spin chain seem to be 'ironed' into much flatter ones compared with those in the parent compound.[14]

Doped holes are primarily located on the spin-chain sublattice in $Sr_{14}Cu_{24}O_{41}$, which fundamentally changes the chain physics. Doped holes combine with $Cu^{2+}$ ions to form nonmagnetic $Cu^{3+}$ ions, becoming so-called Zhang-Rice singlets. Two neighbouring $Cu^{2+}$ spins of each $Cu^{3+}$ ion form a singlet dimer[26] at low temperature (see Fig. 1(b)), in which the two $Cu^{2+}$ spins couple antiferromagnetically.[18] Consequently, such a spin-chain system undergoes into a singlet dimer ground state[18], which is a kind of spin-liquid state with short-range interactions but no long-range magnetic ordering. The chain contribution to the susceptibility shows a broad peak at ~ 80 K,[19] which corresponds to the singlet-triplet excitation of the spin dimers.

Increasing Ca content in $Sr_{14-x}Ca_xCu_{24}O_{41}$ substantially reduces the number of holes on the chains,[16] suppressing the Zhang-Rice singlets and the spin-liquid state.[14] The physics of the spin chains in highly Ca-doped $Sr_{14-x}Ca_xCu_{24}O_{41}$ have not been investigated yet. Considering the large spacings between each chain along both the $a$ and $b$ directions, the chains in $SrCa_{13}Cu_{24}O_{41}$ seems to be an ideal 1D spin system, which becomes an interesting playground for testing low-dimensional quantum effects. By studying the long-range magnetic ordering in $SrCa_{13}Cu_{24}O_{41}$ below 4.2 K, in companion with strong magnetic excitation, we are seeking deep insights to the low-dimensional physics of the spin-chain system in this compound. Therefore, we performed inelastic neutron scattering experiments on a large high-quality single-crystal of $SrCa_{13}Cu_{24}O_{41}$ and determined the spin-wave dispersion and hence the magnetic exchange interaction along the different directions.

**Experiment**

The single-crystal sample of $SrCa_{13}Cu_{24}O_{41}$ for the inelastic neutron scattering experiment was grown by using the travelling-solvent floating-zone technique in an imaging furnace with a high oxygen pressure (~ 32 bar).[27] For the experiment on the time-of-flight (TOF) spectrometer PELICAN[28] at OPAL, the single crystal was aligned in the *bc* plane using by the neutron Laue camera JOEY.[29] An incident neutron beam with a wavelength of 4.69 Å (3.75 meV) was used for this experiment. The energy resolution for this configuration on PELICAN is about 0.135 meV. The sample was rotated over a range of 110$^{o}$ with 1$^{o}$ per step. The experimental data were processed into the scattering function $S(q, w)$ by using the software package HORACE from ISIS.[30] The cold-neutron triple-axis spectrometer SIKA[31] at OPAL was used to measure the spin wave dispersion along the $Q_H$, $Q_K$ and $Q_L$ directions within the *ac* and *bc* scattering planes. A double-focusing monochromator and open-open-60'-60' collimation with a fixed final energy of $E_f$ = 5 meV were used for the experiments on SIKA. A cooled Be-filter was utilized on the scattering side of the instrument in order to eliminate second-order contamination. The energy resolution of this configuration is about 0.13 meV at the elastic line and about 0.16 meV at the energy transfer of 4 meV. The data collected from SIKA were fitted to a double Lorentzian cross-section by convoluting with the instrumental resolution.

**Results and discussions**

The TOF data of $SrCa_{13}Cu_{24}O_{41}$ at 1.5 K are presented in Fig. 2. Fig. 2 (a) and (b) show the energy excitation in an energy window from 0.5 to 2.0 meV in the $Q_K$-$Q_L$ plane and the dispersion along the $Q_L$ direction in the energy range from 0 to 2.5 meV, respectively. These two figures are directly cut from the 3D plot showed in Fig. 2 (c). From the 3D plot, we can clearly identify the inelastic excitation running along all $Q_K$ at $Q_L$ = 3 r.l.u. (reciprocal lattice unit), which corresponds to the horizontal stripe of high intensity at $Q_L$ = 3 r.l.u. in Fig. 2 (a). This stripy excitation should be attributed to its nondispersive feature along the $Q_K$ direction. Viewing in the $Q_L$-E plane in Fig. 2 (b), the apparent broadening of this excitation peak at higher energy transfer indicates its dispersion along the $Q_L$ direction. Further cuts at various energy transfers of Fig. 2 (b) generate the constant energy curves shown in Fig. 2 (d). The fitting to these curves indicate that the excitation peak gradually splits into two peaks when increasing the energy, which confirms the dispersion of this excitation along $Q_L$. All these excitation features disappeared at a raised temperature of 50K (not shown here) except the strong signal observed around (040) (see Fig2 (a)). Therefore, we safely attribute the stripy excitation to the spin wave excitation of the magnetic ordered state, but the excitation near (040) to the phonon excitation. Due to the limited energy-transfer range covered by PELICAN at this configuration, the top of the excitation could not be observed. Within the good energy resolution (~0.135meV) of the current setup, the excitation did not show any energy gap at the zone centre, suggesting a gapless excitation.

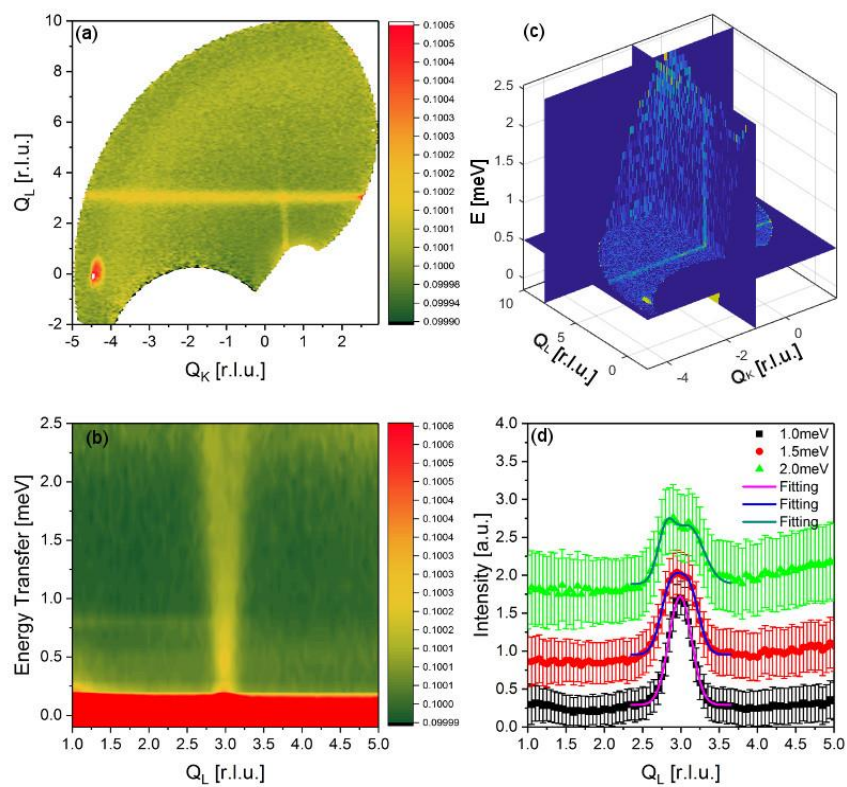

Fig. 2. The low-energy magnetic excitation of SrCa$_{13}$Cu$_{24}$O$_{41}$ measured on PELICAN at 1.5 K. (a) the constant energy cut in the $Q_K$-$Q_L$ plane by integrating intensity over the energy range from 0.5 to 2 meV. (b) The magnetic excitation in the $Q_L$ – E plane with the integrated intensity over the $Q_K$ range from -5 to 0 r.l.u. (c) the 3D view of the magnetic excitation in the $Q_K$ - $Q_L$ - E space. (d) the constant energy cuts at 1meV, 1.5meV and 2meV from (b) with a 1meV window. The curves are intentionally shifted along the intensity axis in order to avoid the overlapping.

In order to measure the magnetic excitation in detail, we carried out inelastic neutron scattering experiments on the cold-neutron triple-axis spectrometer SIKA with a fixed final energy of 5 meV. The results are presented in Fig. 3, 4 and 5. The spin chains in $SrCa_{13}Cu_{24}O_{41}$ run along the $c$ axis of the crystal. The interaction along this direction is supposed to be the strongest and most important for the determination of the dynamic behaviour of the spin chains. Therefore, we present the dispersion of the magnetic excitation along the $c$ axis first. Fig.3 (a) clearly shows an intense excitation peak measured along the $Q_L$ direction with a constant energy transfer of 0.5 meV at 1.5 K. The peak completely disappeared at 100 K, which confirms its magnetic origin. On increasing the energy transfer, the central peak gradually broadens and splits into two peaks while the intensity drops. Above an energy transfer of 2meV, the two peaks are clearly resolved and well separated. This result is consistent with the constant energy cut at 2meV from PELICAN in Fig. 2(d). The asymmetrical shapes of the two peaks are not only due to the instrument resolution, but also due to the inclined background at small scattering angles. The background increases at higher energy transfers. Near the zone boundary, an energy scan was conducted with an extended counting time. Figure 3(f) shows an energy scan at a fixed $Q$-point of (0, 1, 3.9), i.e. close to the boundary of the Brillouin zone. A peak at 5.3 ± 0.2 meV was observed. The data were fitted by convoluting with the instrument resolution.

The spin chains in $SrCa_{13}Cu_{24}O_{41}$ are stacked along the $b$ axis with the alkaline-earth cations and spin-ladder planes in between. We expected that the interaction along the b axis is very weak. In order to confirm this speculation, the measurements at different $Q_K$ positions were carried out on SIKA and the results are shown in Fig. 4. The three $Q_L$ scans at the different $Q_K$ values (=0.8, 1.2, 1.5 r.l.u.) in Fig. 4(a), (b) and (c) basically show similar excitation peaks at the energy transfer of 0.5 meV. Fig. 4(d) shows the comparison of the intensity of the peak at $Q_L$ = 3 r.l.u. and the background intensity at an off-peak position $Q_L$ = 2.5 r.l.u. The intensity of the peak at lower $Q_K$ is slightly higher due to the decrease of the magnetic form factor at higher $Q$ values. The data indicate that there is no dispersion along the $Q_K$ direction. These results agree well with the stripy feature along the $Q_K$ direction observed on PELICAN (see Fig. 2(a) and (c)).

The spin chains lie in the $ac$ plane, in which the interaction between the neighbour chains is another important dynamic factor which determines the dimensionality of this spin-chain system. Our inelastic neutron-scattering experiment on SIKA was performed in the vicinity of $Q$ (0, 0, 3) in the $ac$ scattering plane. Selected results are shown in Fig. 5. Constant-energy scans are plotted in Fig. 5 (a), (b) and (c) for energy transfers of 0.5 meV, 1.5 meV and 2.5 meV, respectively. The peak splitting at the larger energy transfer clearly indicates the dispersion along the $Q_H$ direction. An energy scan was carried out at the boundary of the Brillouin zone at $\mathbf{Q}$ (1, 0, 3) and is shown in Fig. 5(d). The excitation was observed at 4.2 ± 0.2 meV, which is lower than the zone boundary energy along the $Q_L$ direction. All the collected experimental data for the constant-energy and constant-$\mathbf{Q}$ scans for the directions $Q_H$ and $Q_L$ are summarized in the dispersion curve shown in Fig. 6.

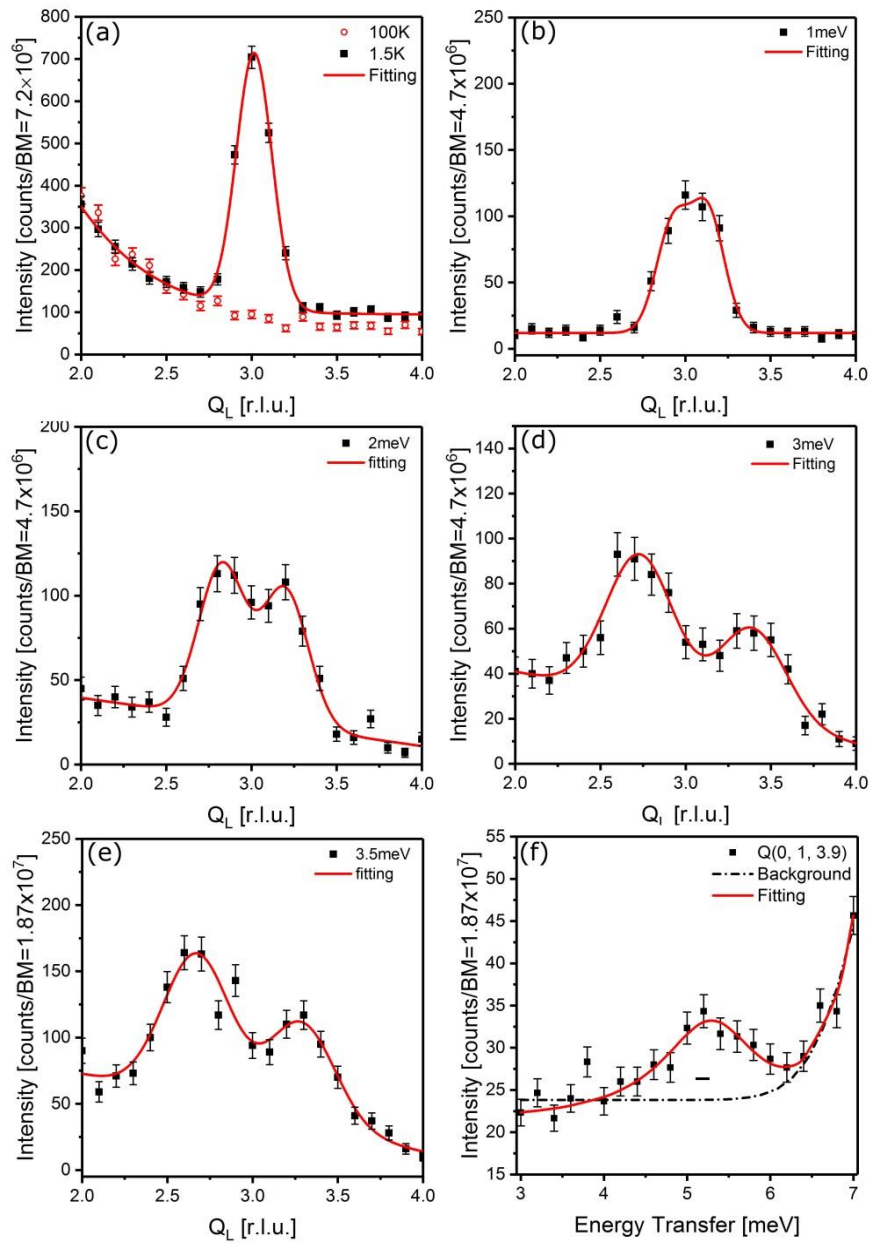

Fig. 3 $Q_L$ scans at energy transfers of 0.5 meV (a), 1 meV (b), 2 meV (c), 3 meV (d) and 3.5 meV (e) through the antiferromagnetic zone centre $\mathbf{Q}$ (0, 1, 3) at 1.5 K with a fixed-$E_f$ mode on SIKA; (f) an energy scan at $\mathbf{Q}$ (0, 1, 3.9), i.e. close to the boundary of the Brillouin zone, with a fixed-$E_f$ mode on SIKA. The solid lines are fitted to the double-Lorentzian cross section convoluted with the instrumental resolution, where the background is shown by the dash-dot line. The horizontal bar below the peak in (f) shows the instrumental resolution at the energy transfer of 4 meV.

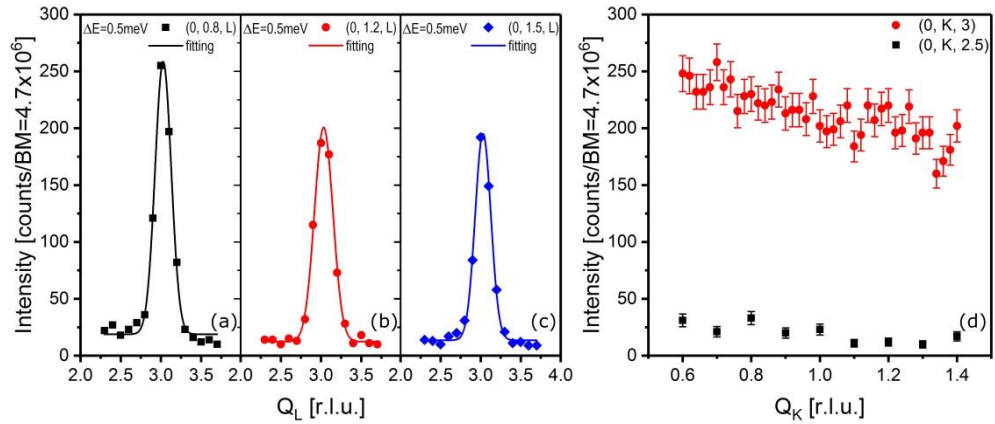

Fig. 4 $Q_L$ scans at the energy transfer 0.5 meV with $Q_K$ = 0.8 (a), $Q_K$ = 1.0 (b), $Q_K$ = 1.2 (c); (d) the $Q_K$ scans at the peak position with $Q_L$=3 and off-peak position at $Q_L$=2.5.

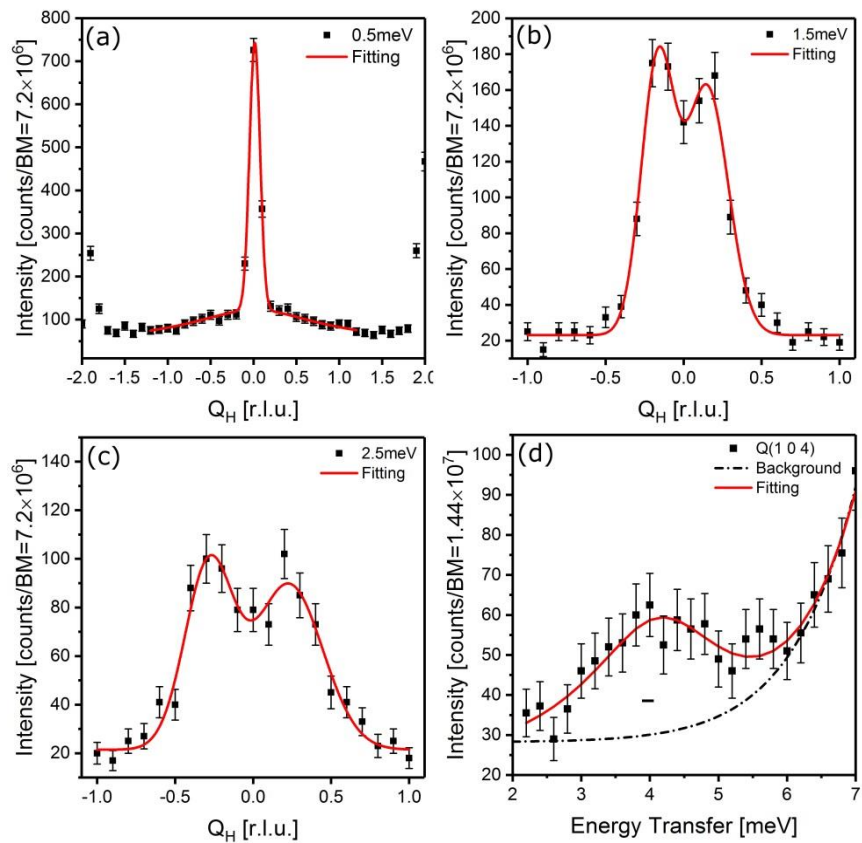

Fig. 5 $Q_k$ scans at energy transfers of 0.5 meV (a), 1.5 meV (b) and 2.5 meV (c) through $\boldsymbol{Q}$ (0, 0, 3) with a fixed-$E_f$ mode on SIKA; (d) an energy scan at the boundary of the Brillouin zone at $\boldsymbol{Q}$ (1, 0, 3) with a fixed-$E_f$ mode on SIKA. The solid lines are fitted to the double-Lorentzian cross section convoluted with the instrumental resolution. The background is shown by the dash-dot line. The horizontal bar below the excitation peak in (d) shows the instrumental resolution at the energy transfer of 4 meV.

Both the spin-ladder and spin-chain subsystems in SrCa$_{13}$Cu$_{24}$O$_{41}$ are quasi-1D or 1D magnetic systems. Our previous inelastic neutron-scattering experiment confirmed the spin-gap excitation at ~ 32 meV in this compound.[16, 17] This excitation was attributed to the spin ladder due to its consistency with most other results reported for the spin ladders in the parent compound.[22, 32] The exchange interactions along the legs and rungs were estimated to be ~ 130 meV and ~ 72 meV,[22] respectively, or larger, which agrees well with the antiferromagnetic interactions through the 180$^o$ Cu-O-Cu pathway in cuprates with square lattices[33]. Therefore, we attribute the low-energy excitation observed above to the spin-chain sublattice. As reported in other edge-sharing CuO$_2$ chain systems, such as CuGeO$_3$,[34] the exchange interactions between neighbouring Cu$^{2+}$ ions through a 90$^o$ Cu-O-Cu pathway are much weaker than the excitation in the corner-sharing CuO$_2$ systems with an 180$^o$ Cu-O-Cu pathway.

As described above and shown in Fig. 1, the spin chains in SrCa$_{13}$Cu$_{24}$O$_{41}$ are built by edge-sharing CuO$_2$ squares. Comparing with a corner-sharing spin chain, an edge-sharing spin chain has much weaker interaction ($J_c$) within the chains. Along the stacking direction, namely, the $b$ axis, the interaction is very weak and negligible due to the large interlayer spacing, which is confirmed by the dispersionless behaviour of the spin excitation as shown above. Similar weak interlayer interactions have been reported in La$_2$CuO$_4$[33] and Rb$_2$MnF$_4$,[35] both of which were classified as 2D magnets. Considering the intralayer interactions, there are several possible interactions, such as $J_c$, $J_{ac1}$, $J_{ac2}$ and $J_a$, as shown in Fig. 1(c). Ignoring the single-ion anisotropy and taking in-plane intrachain and interchain interactions into account, the Heisenberg Hamiltonian of the spin-chain system in SrCa$_{13}$Cu$_{24}$O$_{41}$ reads as follows:

$$H = \sum_{i,\,j} J_{ij} \vec{\boldsymbol{S}}_i \cdot \vec{\boldsymbol{S}}_j \qquad \text{Eq. (1)}$$

where $J_{ij}$ includes $J_c$, $J_{ac1}$, $J_{ac2}$ and $J_a$ in Fig. 1(c). According to LSWT, the dispersion relationship could be derived into the following formula[36]:

$$h\omega = S\left\{[J_c(\cos(q_L)-1)+J_a(\cos(q_H)-1)+2J_{ac1}+2J_{ac2}]^2\right.$$

$$\left.-\left[2J_{ac1}\cos\left(\frac{q_H}{2}\right)\cos\left(\frac{q_L}{2}\right)+2J_{ac2}\cos\left(\frac{q_H}{2}\right)\cos\left(\frac{3q_L}{2}\right)\right]^2\right\}^{1/2}$$

<div align="right">Eq. (2)</div>

where $S$ is the moment size, $q_H$ and $q_L$ correspond to the reciprocal lattice units along the a* and c* directions.

We have tried many different combinations of the different exchange interactions in the model above to fit the experimental data. Finally, it was found that the experimental data cannot be well described if $J_{ac1}$ and $J_{ac2}$ are in the comparable range as $J_c$ and $J_a$. By ignoring these two exchange interactions from the model above, therefore, we are able to fit our data well to the model, which gives $J_c$=5.4 ± 0.1meV and $J_a$=4.4 ± 0.3 meV. The NN intrachain interaction is antiferromagnetic, which was confirmed by the linear relationship of the dispersion curve observed on both PELICAN and SIKA at the low $q_L$ range. Fig. 6 (a) and (b) shows the fitting to the experimental data along the $Q_L$ and $Q_H$ directions with the LSWT model. The fitted dispersion relations are quite consistent with the experimental results.

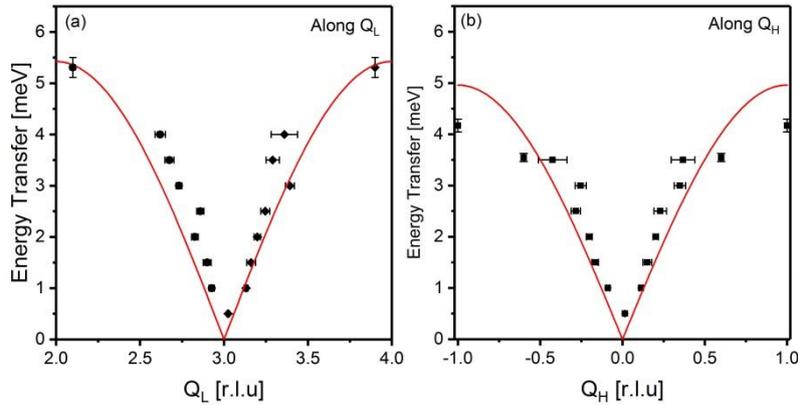

Fig. 6 The dispersion curves along the $Q_L$ (a) and $Q_H$ (b) directions. The black dots are experimental data while the red curves are the dispersion relations as obtained by fitting to the right branches of the experimental data while the left branches are plotted symmetrically with the same fitted parameters.

In comparison with the 2D magnetic systems, such as La$_2$CuO$_4$[33] and Rb$_2$MnO$_4$,[35] the spin chain system in SrCa$_{13}$Cu$_{24}$O$_{41}$ has the similar 2D spin wave feature. According to the literature, the interplane interaction is several magnitude orders weaker than the intraplane interaction in La$_2$CuO$_4$ and Rb$_2$MnO$_4$, which causes the nondispersive spin wave excitation along the stacking direction of layers. However, such a relatively weak interaction is strong enough to drive these two systems into a long range magnetic ordering states. For example, $Rb_2MnO_4$ orders into the antiferromagnetic phase at 38.4K.[35] In contrast, $SrCa_{13}Cu_2O_{41}$ becomes antiferromagnetic ordering at 4.2K[17], indicating a much weaker interplane interaction.

It is interesting to compare the current spin-chain system with other edge-sharing spin chains, such as $CuGeO_3$[11, 34], $Ca_2Y_2Cu_5O_{10}$[12, 36], and $Li_2CuO_2$[13, 37]. $CuGeO_3$ has similar spin chains as the current spin-chain system and undergoes a spin-Peierls phase transition at 14 K, where the $Cu^{2+}$ spins form a spin-dimer ground state. Inelastic neutron scattering reveals that the nearest-neighbour exchange $J_c$ is antiferromagnetic and around 10.4meV. The nearest intrachain interaction of $Y_2Ca_2Cu_5O_{10}$ is ferromagnetic ( ~ -8 meV). The chain in $Li_2CuO_2$ is also ferromagnetic with an nearest intrachain interaction of ~ -14.7 meV, reported by a recent study[13]. In the previous study on the similar compound, $La_5Ca_9Cu_24O_{41}$,[9] it was found that its NN exchange interaction is weak ferromagnetic (-0.2 meV). However, the magnetic phase orders at about 10K, which is not consistent with the weak antiferromagnetic interaction. From the comparison of all these studies, we conclude that the NN intrachain interactions of these chain systems vary dramatically. The variation of the Cu-O-Cu bond angles and lengths could be the main reason. According to the Goodenough-Kanamori-Anderson rule,[38] when the Cu-O-Cu bond angle changes from $180^o$ towards $90^o$, the exchange interaction evolves from a positive (antiferromagnetic) value to a negative (ferromagnetic) value. According to our previous study,[14] the variation of the bond angle Cu-O-Cu significantly increases with the increase of Ca doping, in which the maximum angle could reach $96^o$ due to the structural modulation of this incommensurate system. Such variation may be even larger in $SrCa_{13}Cu_{24}O_{41}$. Such a bond-angle variation could be the main reason for the antiferromagnetic interaction observed in this compound, which is evidenced by the sharp linear dispersion at the low $q_L$ range seen in the PELICAN data.

Since the parent compound $Sr_{14}Cu_{24}O_{41}$ and $SrCa_{13}Cu_{24}O_{41}$ have very similar crystal structures, one may expect that they should have similar dynamic behaviour from their spin-chain sublattices. Actually, the previous and current experimental results indicate that this is not the case. Matsuda et al.[18] have measured the magnetic excitation from the spin-chain sublattice of $Sr_{14}Cu_{24}O_{41}$. They observed that the excitation ranges from 9 meV to 11 meV at low temperature, showing very little dispersion along the chain. This observation was explained by the spin-dimer model for the chains. As mentioned above, the spin-chain sublattice in $Sr_{14}Cu_{24}O_{41}$ is highly doped with holes. There are about 5.5 holes on 10 $CuO_2$ square units. At low temperature, these holes localize to $Cu^{2+}$ ions to form Zhang-Rice singlets.[26] The two neighbouring $Cu^{2+}$ ions of each Zhang-Rice singlet form a spin $\downarrow$ and spin ↑ singlet dimer state. Due to the large number of the holes on the chains, these singlet dimers dominate the physics of the spin-chain sublattice of $Sr_{14}Cu_{24}O_{41}$ at low temperature. The magnetic excitation observed at ~ 10 meV corresponds to the singlet-to-triplet excitation of these spin dimers, which explains the spin gap and the broad peak at ~ 80 K in the susceptibility measurement[39].

The excitation scenario is completely different in $SrCa_{13}Cu_{24}O_{41}$. Due to the high Ca doping level, the holes are mainly driven to the spin-ladder sublattice in $SrCa_{13}Cu_{24}O_{41}$.[14] The spin chain in this compound, therefore, becomes a "non-hole-doped" $Cu^{2+}O_2$ spin chain. Due to the magnetic ordering, the excitation from the spin chains in this compound is either a spin-wave type or a spinon-type of excitation. The above analysis of the inelastic neutron-scattering data has confirmed the 2D spin-wave excitation in $SrCa_{13}Cu_{24}O_{41}$. The major difference of the chain dynamics in $SrCa_{13}Cu_{24}O_{41}$ and $Sr_{14}Cu_{24}O_{41}$ is that the former has a gapless excitation while the latter has a gapped excitation. The former one propagates in the magnetic lattice of the spin chain, especially along the chain, while the latter is nearly localized.

In conclusion, we studied the low-energy excitation spectrum from the magnetic ordered state of $SrCa_{13}Cu_{24}O_{41}$. The excitation shows dispersions along the *c* and *a* directions but is nondispersive along the *b* direction, indicating a 2D magnetic nature. The excitations were confirmed to originate from the spin-chain sublattice in this compound. Fitting to the experimental data to LSWT, the nearest-neighbour interaction of the chain was found to be 5.4 meV and antiferromagnetic, which is comparable with the value observed in $CuGeO_3$. The current magnet is different from the spin chain in the parent compound $Sr_{14}Cu_{24}O_{41}$ because the chain physics of $Sr_{14}Cu_{24}O_{41}$ is dominated by the spin dimers around Zhang-Rice singlets. The spin chains of $SrCa_{13}Cu_{24}O_{41}$ are continuous and free of doped holes due to holes transferring to the ladder sublattice. This fact explains the spin gap excitation in $Sr_{14}Cu_{24}O_{41}$ but the gapless excitation for $SrCa_{13}Cu_{24}O_{41}$ in this study. The interaction between the spin chain and ladder systems in the magnetic ordered state of this magnet is still an open question for the further study in the future.

## Acknowledgement

The authors thank ANSTO for the allocation of neutron beam time on SIKA and PELICAN (P5715, P5214, and P4335). G.D. and H.L. thank the support from the National Natural Science Foundation of China (Grant No. 11674372). H. L. and S. L. are also supported by the National Key Research and Development Program of China (Grants No. 2017YFA0302903, No. 2017YFA0303103, No.

2016YFA0300502), the Strategic Priority Research Program (B) of the Chinese Academy of Sciences (Grant No. XDB07020300), and the Youth Innovation Promotion Association of Chinese Academy of Sciences (No.2016004). W.R. and S.C. thank the support from the National Natural Science Foundation of China (NSFC, Nos. 11774217, 51372149, 51672171), the National Key Basic Research Program of China (Grant No. 2015CB921600).